# Electron-like high-temperature superconductivity induced by compressive strain in $La_2PrNi_2O_7$ thin films

Zhiwei Wang[1], Zhengjie Wang[1], Huiyu Wang[1], Mingyi Zhu[1], Mengzu Shi[2], Zongyao Huang[1], Houpu Li[1], Shoucong Ning[1], Jing Tao[1], Tao Wu[1,2,3,4*], and Xianhui Chen[1,2,3,4*]

[1] Department of Physics, University of Science and Technology of China, Hefei 230026, China
[2] Hefei National Research Center for Physical Sciences at the Microscale, University of Science and Technology of China, Hefei 230026, China
[3] Collaborative Innovation Center of Advanced Microstructures, Nanjing University, Nanjing 210093, China
[4] Hefei National Laboratory, University of Science and Technology of China, Hefei 230088, China.
*Correspondence to: wutao@ustc.edu.cn, chenxh@ustc.edu.cn

**The realization of high-temperature superconductivity in bilayer nickelates under epitaxial compressive strain is widely interpreted as mimicking the effects of high hydrostatic pressure. To test the equivalence of these mechanisms, we investigated a comprehensive strain continuum ranging from compressive (−2.14%) to tensile (+0.91%). Crucially, via ozone-assisted atomic-layer epitaxy, we realized high-temperature superconductivity in as-grown $La_2PrNi_2O_7$ films on $NdAlO_3$ substrates, which induce the most extreme compressive strain in this material system. Under extreme compression (−2.14%), these films exhibit a $T_{c,onset}$ of 60 K, zero resistance at 33 K, and a diamagnetic response at 20 K, with magnetotransport measurements confirming a quasi-two-dimensional superconducting nature. Comparing our phase diagram with reported data reveals distinct lattice responses: unlike in pressurized crystals, the superconducting window in epitaxial films diverges significantly in the out-of-plane parameter $c$ (or $c/a_p$ ratio) but remains consistent with the bulk regarding the in-plane parameter $a_p$. Crucially, while superconductivity in both systems emerges from the suppression of spin-density waves (SDW), Hall measurements reveal a fundamental electronic dichotomy: optimal superconducting films are intrinsically electron-like (exhibiting a negative Hall coefficient), in stark contrast to the hole-like nature (positive Hall coefficient) of high-pressure bulk crystals and non-superconducting tensile films. Ultimately, both tuning**

**strategies effectively modulate the underlying correlation landscape—the true driver of superconductivity—transcending the constraints of specific Fermi surface topologies. This work establishes a macroscopic platform for probing the multi-orbital physics of nickelates, offering a new dimension for investigating high-temperature superconductivity.**

## Introduction

Since the discovery of superconductivity in infinite-layer $NdNiO_2$[1], nickelates have emerged as the new family of unconventional high-$T_c$ superconductors[2], following the cuprates and iron-based superconductors[3,4]. Especially, discoveries of superconductivity with $T_c$ above 80 K in pressurized Ruddlesden-Popper (RP) phase nickelates (including bilayer, trilayer, and hybrid phases) have sparked intense interest in exploring their origin and mechanism[5–8]. Unlike their infinite-layer counterparts[2], bilayer nickelates have a nominal $3d^{7.5}$ configuration and retain apical oxygen atoms between adjacent $NiO_2$ planes, suggesting that their pairing may involve a multiband mechanism involving both the $d_{x^2-y^2}$ and $d_{z^2}$ orbitals[5,9,10]. Early studies attributed bulk superconductivity to the pressure-driven orthorhombic-to-tetragonal transition and enhanced interlayer $d_{z^2}$ coupling[5,9]. Emerging evidences, however, suggest that the suppression of competing spin density wave (SDW) order is the fundamental prerequisite[7,11,12].

Experimentally, ultrahigh-pressure conditions impede high-precision spectroscopy and complicate transport measurements. Intrinsic properties such as pairing symmetry thus remain controversial[13,14], obscuring a clear physical interpretation. The recent breakthrough of bilayer nickelate thin films shows that biaxial compressive strain mimics hydrostatic pressure[15–19]: −2.01% strain on $SrLaAlO_4$ (SLAO) induces ambient-pressure superconductivity in $La_3Ni_2O_7$ (LNO) / $La_2PrNi_2O_7$ (LPNO) films ($T_{c,onset}$ ranging from 26 K to above 60 K), whereas weaker strain (−1.07%) on $LaAlO_3$ (LAO) suppresses $T_{c,onset}$ to nearly 10 K. However, a structural paradox remains: while hydrostatic pressure compresses the lattice globally, in-plane epitaxial compression induces a substantial $c$-axis expansion via the Poisson effect. This geometric divergence and the $T_c$ discrepancy relative to pressurized bulk raise the question: does the superconductivity in strained films and high-pressure bulk share the same superconducting conditions? Furthermore, unlike the linear-resistivity behavior in optimal-pressure bulk superconductors[8,20,21], strained films exhibit normal-state transport ranging from Fermi-liquid to non-Fermi-liquid[18,19]. This difference suggests that, while strain induces ambient-pressure superconductivity in thin films, it remains distinct from its high-pressure bulk counterpart—a difference that merits further investigation.

## Epitaxial strategy and strain engineering of $La_2PrNi_2O_7$ thin films

Accessing the intrinsic physics of bilayer RP nickelates thin films is often impeded by their thermodynamic metastability[15,18]. Conventional synthesis routes face significant hurdles: pulsed laser deposition (PLD) frequently suffers from cation off-stoichiometry and impurity phases[22,23], while molecular beam epitaxy (MBE) typically necessitates harsh post-growth annealing to rectify oxygen deficiency[24]. To circumvent these challenges, we stabilized the bilayer phase and suppressed oxygen vacancies by partially substituting Pr for La (yielding $La_2PrNi_2O_7$)[8]. We further optimized an ozone-assisted atomic-layer-by-layer epitaxy technique[25–28] (Extended Data Fig. 1; see Methods), where high-purity ozone ensures ideal oxygen stoichiometry in as-grown films, thereby eliminating the need for post-annealing. To minimize interfacial energy[29], we engineered a specific monolayer buffer (Fig. 1a) guided by established natural interfacial configurations[17,27,30,31]. Furthermore, to accommodate the dynamic atomic rearrangement inherent to RP-phase growth[32–34], we introduced a brief in situ annealing step after each atomic layer (Extended Data Fig. 2), ensuring superior crystalline quality.

Leveraging this robust synthesis strategy, we established a comprehensive strain continuum ranging from extreme compression (−2.14%) to tension (+0.91%) using four distinct substrates: $NdAlO_3$ (NAO), SLAO, LAO, and $(La,Sr)(Al,Ta)O_3$ (LSAT). Crucially, we introduce the Sr-free NAO substrate for the first time. NAO not only imparts extreme compressive strain (-2.14%) but also eliminates the interfacial Sr doping prevalent in SLAO[15,16]. While moderate Sr doping may aid in stabilizing superconductivity[24,35], a critical question remains: is the ambient-pressure superconductivity driven by intrinsic in-plane compressive strain or parasitic interfacial doping? Therefore, we first focused on NAO substrates to explore superconductivity in extremely strained nickelate thin films.

## Superconducting properties of $La_2PrNi_2O_7$ thin films on NAO substrates

Employing the aforementioned growth strategy with the ozone-assisted atomic-layer-by-layer epitaxy technique, we grew phase-pure three-unit-cell-thick (3UC) LPNO films on NAO substrates (Fig.1a). As presented in Fig. 1b, the in-situ reflection high-energy electron diffraction (RHEED) intensity oscillations afforded real-time tracking of individual atomic layers with distinct chemical compositions. With the exception of the initial (La,Pr)O layer—whose V-shaped oscillation profile likely originates from the charge-mismatch reconstruction between the $AlO_2$ termination and the first (La,Pr)O layer—the subsequent single-layer growth strictly obeys the standard pattern: intensity attenuation during rare-earth oxide layer deposition and recovery during $NiO_2$ layer deposition. This RHEED evolution is consistent with previous reports on high-quality films and is driven by the combined effects of surface roughness kinetics and mean inner potential[16,24,36]. The continuous, unattenuated RHEED intensity oscillations during growth, along with the sharp post-growth diffraction streaks shown in Extended Data Fig. 4a, demonstrate the atomic layer-by-layer growth of our films.

Having established this extreme-strain platform, we investigated its electrical transport properties. Fig. 1c presents the transport behavior of a representative sample (Sample A). By analyzing the first derivative of resistivity with respect to temperature ($d\rho/dT$), plotted in Fig. 1d, we precisely extracted a superconducting onset transition temperature ($T_{c,onset}$) up to 60 K, which coincides with the onset of the 9 T magnetic field response (Fig. 1d, Extended Data Fig. 3a). On further cooling to 33 K, the film resistance drops below the instrumental resolution limit, and enters a complete zero-resistance state ($T_{c,zero}$, shown in Fig. 1e and Extended Data Fig. 3b). This result provides definitive proof that intrinsic in-plane biaxial compressive strain—decoupled from parasitic interfacial hole doping—is alone sufficient to induce high-temperature superconductivity approaching 60 K in strained bilayer films. Remarkably, prior to the superconducting transition, the film exhibits strong metallicity over a wide temperature range, accompanied by signatures of resistivity saturation at higher temperatures. A power-law fit ($\rho(T) = \rho_0 + AT^n$) to the 60–80 K range yields $n$ = 1.95, consistent with literature reports[18,37], but the fit deteriorates at higher temperatures. We therefore apply the Mott-Ioffe-Regel (MIR) limit fitting method used in previous nickelate films studies[17,24] to the normal-state resistivity (see Methods for details), yielding a phenomenological scattering exponent $n$ ≈ 2.6. These fits indicate that the normal state of $La_2PrNi_2O_7$/NAO thin film does not exhibit the $T$-linear resistivity observed in high-pressure bulk samples[8,20,21].

To confirm the intrinsic, bulk nature of the superconductivity in our films, we utilized a two-coil mutual-inductance technique[38] (see Methods) to measure the AC diamagnetic susceptibility response (Fig. 1f) for Sample A. The experimental data reveal the onset of a macroscopic Meissner diamagnetic state at approximately 20 K, accompanied by a pronounced drop in the real part of the voltage and a characteristic dissipation peak in the imaginary part at lower temperatures. As shown in Extended Data Fig. 3c, an additional sample (Sample B) also displayed a Meissner transition comparable to that previously reported[18,19]. These results further confirm that the bulk superconductivity is stabilized by biaxial compressive strain in LPNO thin films. Rather than merely serving as an additional substrate, this doping-free LPNO/NAO system establishes an ideal extreme-compression benchmark to explore the underlying physics of ambient-pressure nickelate superconductivity.

**Epitaxial strain and atomic-scale structural integrity of $La_2PrNi_2O_7$/NAO thin films**

To evaluate the crystalline quality and strain state of the LPNO/NAO films, we performed systematic macroscopic structural characterization. As shown in Fig. 2a, the X-ray diffraction (XRD) $2\theta$-$\omega$ scan along the (001) direction reveals distinct (00$l$) peaks at their expected positions, with no evidence of intergrown RP impurity phases (e.g., $n$ = 1 or 3). Notably, the intense and sharp (004) peak, accompanied by pronounced Laue oscillations, attests to the high crystalline quality of the film. Furthermore, sharp in situ diffraction spot patterns corroborate atomic force microscopy (AFM) results (Extended Data Fig. 4a,b), confirming an atomically smooth surface and interface. Reciprocal space mapping (RSM) around the NAO (103) and LPNO ($10\underline{17}$) reflections (Fig. 2b) confirms that the film is fully coherently strained to the substrate, with its in-plane lattice constant locked to the substrate value (3.751 Å). The $c$-axis lattice constant, determined via a Nelson–Riley fit to $2\theta$–$\omega$ scans, is 20.86 Å, corresponding to an elongation of ~1.66% relative to the bulk material[5]. Consistent with this, X-ray reflectivity (XRR) yields a 6.3 nm thickness (Extended Data Fig. 4c)—precisely matching the 3UC film with buffer layer and validating our atomic-layer epitaxy.

To verify crystallinity and interfacial configuration at the atomic scale, we conducted large-area high-angle annular dark-field scanning transmission electron microscopy (HAADF-STEM). Over a 50-nm field of view (Fig. 2c), the film exhibits a continuous RP bilayer structure, ruling out stacking faults or intergrown impurities. High-resolution imaging further reveals an interfacial reconstruction matching our design—a complete $AlO_2$–(La,Pr)O–$NiO_2$–(La,Pr)O buffer on the $AlO_2$-terminated NAO substrate—followed by the standard (La,Pr)O–$NiO_2$–(La,Pr)O–$NiO_2$–(La,Pr)O film stacking. The deviation of this entire configuration from the programmed deposition sequence (see Methods) demonstrates dynamic atomic rearrangement during growth. Correspondingly, large-area annular bright-field (ABF) images display a consistent structural profile (Extended Data Fig. 5). Additionally, atomically resolved energy-dispersive X-ray spectroscopy (EDS) elemental mapping (Fig. 2d) precisely delineates the spatial distribution of each chemical species, confirming the absence of detectable ionic interdiffusion.

## Quasi-2D superconductivity in $La_2PrNi_2O_7$/NAO thin films

To further elucidate the dimensionality of the superconducting state in this purely strained system, we performed systematic transport measurements on the LPNO/NAO films under applied magnetic fields of up to 9 T. When the magnetic field is applied either perpendicular to ($H \perp ab$) or parallel to ($H \parallel ab$) the *ab*-plane, the superconducting transition is progressively suppressed, revealing a highly pronounced anisotropic response, as shown in Fig. 3a,b. Notably, $H_{c2}^{\perp ab}(T)$ displays a typical linear temperature dependence, whereas $H_{c2}^{\parallel ab}(T)$ exhibits a distinct non-linear square-root dependence. By extrapolating these temperature-dependent behaviors (using the 90% normal-state resistance criterion), we obtained remarkably high zero-temperature upper critical fields of $H_{c2}^{\perp ab}(0) \approx 84.1\mathrm{T}$ and $H_{c2}^{\parallel ab}(0) \approx 144.2\mathrm{T}$ (Fig. 3c; see Methods for detailed fitting). Although the extracted anisotropy factor $\gamma$ is approximately 1.71—a value superficially implying a degree of 3D coupling—the stark dichotomy in the temperature dependence unambiguously demonstrates its 2D superconducting nature. This behaviour is fundamentally distinct from the 3D characteristics observed in infinite-layer nickelates[39,40].

Using the standard Ginzburg-Landau (GL) formalisms (see Methods), we extract zero-temperature in-plane coherence length $\xi_{ab}(0)$ is approximately 1.98 nm, highly consistent with previously reported values[16,18,24,41]. Crucially, the calculated effective superconducting layer thickness ($d \approx 4$ nm) is remarkably close to the physical film thickness of ~6 nm. This agreement once again excludes the possibility of parasitic interfacial superconductivity, confirming that the superconducting state is intrinsic and driven purely by biaxial strain, not interfacial doping.

To rigorously define the superconducting dimensionality, we examined the angular evolution of $T_{c,50\%}$ under a constant magnetic field of 9 T (Fig. 3d). The experimental data reveal an extremely sharp cusp-like feature near $\theta = 0$ degrees ($H \parallel ab$), which is perfectly captured by the 2D Tinkham model (solid line in Fig. 3d; see Methods for detailed fitting). In stark contrast, the 3D anisotropic GL model completely fails to reproduce this angular response (dashed line). Together, these critical field analyses firmly establish the quasi-2D nature of this doping-free LPNO/NAO film system near the transition temperature. However, given that previous ambient-pressure nickelate films often exhibit stronger three-dimensional superconducting characteristics in the zero-temperature limit under high magnetic fields[37,42], the ultimate dimensionality of the superconducting state in our LPNO/NAO films at extreme limits remains an open question that merits further high-field transport measurements.

**Hall coefficients in $La_2PrNi_2O_7$ thin films with different substrates**

To further probe the evolution of the normal-state carrier characteristics in LPNO/NAO system, we performed Hall resistivity measurements from 50 K to 200 K, as presented in Extended Data Fig. 8a. Although the electronic structure of ambient-pressure superconducting nickelate thin films exhibits complex multiband itinerant characteristics at the Fermi level[43,44], our results demonstrate that the Hall resistivity varies strictly linearly with the magnetic field at all measured temperatures. Furthermore, the Hall coefficient ($R_H$, Fig. 3e) evolves smoothly and remains consistently negative within the measured temperature range. The absolute $R_H$ of our LPNO/NAO film exceeds those of previously reported optimized films[16,18,24,41]. Meanwhile, this consistently negative behavior closely parallels theoretical calculations[45] and experimental observations on SLAO substrates, including both our own LPNO films and these counterparts. However, it diverges from superconducting pressurized bulk crystals, which exhibit a positive, significantly smaller normal-state $R_H$[20,46]. Previously, this sign divergence is always attributed to distinct Fermi surface topologies between the two platforms, particularly the relative position of the $\gamma$ band[5]. Here, we think that this structural (Poisson effect) and electronic (Hall sign) disparity raises a critical question: how does superconductivity universally emerge across such distinct states? To resolve this, we expand our investigation from this extreme compressive benchmark to a broad epitaxial strain continuum.

Using this similar epitaxy strategy (see Methods), we synthesized 3UC LPNO films on three additional substrates with varying lattice constants. As shown in Fig. 4a, the four selected substrates—NAO, SLAO, LAO, and LSAT—encompass a broad epitaxial strain range from -2.14% (strong in-plane compression) to +0.91% (in-plane tension) relative to the bulk material[5]. Through $2\theta$-$\omega$ scans along the (001) direction and reciprocal space mapping (RSM) around the substrates (103) / $(10\underline{11})$ and films $(10\underline{17})$ reflections (Extended Data Figs. 6,7), we precisely extracted the in-plane ($a_p$) and out-of-plane ($c$) lattice constants of the films under each strain state (Extended Data Table. 1). The corresponding transport evolution is presented in Fig. 4b, displaying typical $\rho$–$T$ curves for LPNO films across entire strain continuum.

We subsequently tracked the normal-state evolution under strain via temperature-dependent Hall measurements. As detailed in Extended Data Fig. 8, the Hall resistivity measured across all strain states exhibits a strictly linear dependence on the applied magnetic field over the entire temperature range. From this linearity, we extracted the continuous evolution of the $R_H$, utilizing the ambient-pressure, non-superconducting bulk as a critical reference[47] (Fig. 4c). As discussed earlier, for the ambient superconducting films grown on highly compressive NAO and SLAO, $R_H$ remains negative, exhibiting a systematic temperature-dependent shift as the compressive strain relaxes. In contrast, films grown under weak compression or tension (LAO and LSAT) exhibit a positive $R_H$. This behavior closely resembles to that of ambient-pressure bulk samples[47], where a SDW may partially gap the electron pocket and leave hole-like bands to dominate transport. Additionally, to phenomenologically track the dominant scattering mechanisms, we extracted the scattering exponent $n$ from MIR saturation fits across all four substrates, selecting five representative high-quality samples for each (Extended Data Fig. 9, Extended Data Table 2; see Methods). The extracted $n$-values decrease monotonically from extreme compression to tension, revealing a profound strain-driven modulation of the normal-state scattering mechanism. However, since these $n$-values represent an average over multiple scattering channels across a wide temperature range, a definitive quantitative analysis of these mechanisms will likely require further studies.

## Correlation between $T_c$ and lattice parameters in $La_2PrNi_2O_7$ thin films

Ultimately, we constructed a structural phase diagram for the bilayer nickelate system by correlating our $T_{c,onset}$ with the extracted lattice parameters, integrating previous reports on high-pressure bulk crystals and optimized $(La,Pr,Sr)_3Ni_2O_7$ films [5,8,16–19,24,48,49] (Fig. 4d–f). As shown in Fig. 4d, films and bulk crystals share a consistent in-plane trajectory: extreme in-plane compression (reduced $a_p$) is universally critical for driving the rapid enhancement of $T_{c,onset}$. This establishes a universal in-plane parameter regime ($a_p < 3.79$ Å) within which robust high-$T_c$ superconductivity emerges, irrespective of whether the lattice compression is driven by hydrostatic pressure or epitaxial strain.

However, the out-of-plane parameter $c$ and the $c/a_p$ ratio reveal a stark dichotomy. Under hydrostatic pressure, superconductivity in bulk crystals emerges with a reduced $c$-axis parameter ($c$ < 20.14 Å) and a deeply contracted regime ($c/a_p$ < 5.34) (Fig. 4e, f). In contrast, under epitaxial strain, superconductivity in thin films occurs with an elongated $c$-axis ($c$ > 20.47 Å) and a relatively expanded regime ($c/a_p$ > 5.40). These disparities in out-of-plane structural parameters underscore a fundamental divergence in the superconducting conditions between strained films and pressurized bulk crystals, providing a structural origin for the opposite Hall coefficients discussed above. Notably, recent DFT calculations[45] suggest that conventional Fermi surface topology alone cannot explain these contrasting Hall coefficients, highlighting the need for further theoretical investigation into the role of electronic correlation effects.

To elucidate this distinct macroscopic trend, we integrate our transport phase diagram with recent microscopic insights[45,50,51]. Fundamentally, both epitaxial strain and hydrostatic pressure induce superconductivity by suppressing a competing SDW[11,12,51]. Despite the differences in structure and Fermi surface between the superconducting films and their high-pressure bulk counterparts (as discussed before), microscopically, both strain and high pressure straighten Ni-O-Ni bond angles and suppress local Jahn-Teller octahedral distortions[45,52]. This overarching structural symmetry restoration triggers a Ni $3d_{z^2}$ orbital delocalization and a profound "magnetic collapse" [51]. However, their out-of-plane dynamics starkly diverge. While bulk hydrostatic pressure directly compresses the $c$-axis to enhance interlayer hopping ($t_\perp$), our epitaxial films experience a Poisson-induced $c$-axis expansion[53]. Addressing this discrepancy, resonant inelastic x-ray scattering (RIXS) studies reveal that extreme in-plane confinement actually broadens the spin-excitation bandwidths[50], which strengthens the interlayer magnetic exchange coupling ($J_\perp$) despite the expanded $c$-axis[50,54]. Ultimately, extreme biaxial strain not only mimics the magnetic suppression of high pressure but also creates a uniquely coherent structural and electronic regime inherently optimized for robust 60 K superconductivity in nickelate films.

## Discussion and conclusion

Recent studies report normal-state $T$-linear resistivity in bulk crystals near optimal pressure[8,20,21]. However, across our entire strain continuum, superconductivity consistently emerges from a metallic normal state lacking this signature, as observed in the high-$T_c$ (~60 K) films on both NAO (Fig. 1) and SLAO substrates (Extended Data Fig. 10). Given the controversial normal-state behaviors reported in recent literatures[19,37,55], our findings indicate that $T$-linear strange-metal transport is not a prerequisite for high-$T_c$ superconductivity in nickelate films, an observation that warrants further investigation.

Unlike single-band cuprates where the carrier doping is a necessary prerequisite for achieving high-$T_c$ superconductivity[3], bilayer nickelates represent a distinct class of multi-orbital correlated metals[9,56–58], in which carrier doping is not a prerequisite. The opposite Hall behaviors (negative vs. positive $R_H$) between the strained superconducting films and pressurized superconducting bulk crystals should not be attributed solely to differences in Fermi surface topology. Instead, extreme in-plane compressive strain might primarily modulate the underlying electronic correlations. The systematic evolution of both the normal-state scattering behavior and the sign of $R_H$ likely reflects the universal suppression of the competing SDW state and a fundamental reorganization of orbital-dependent correlation effect, transcending the specific topological differences of the Fermi surface. While recent high-pressure experiments on epitaxial films[59–61] suggest that substrate-induced structural penalties fundamentally limit the ambient-pressure $T_c$, our findings reveal a different pathway: extreme in-plane compression may help mitigate these limitations via the modulation of electronic correlations. Building on this momentum, advancing epitaxy to stabilize films on even more compressive substrates, such as $NdCaAlO_4$ (001) and $YAlO_3$ (110), could further optimize these orbital-dependent interactions and pairing strength[62], potentially elevating the ambient-pressure thin-film $T_c$ to approach optimal bulk values. Crucially, by anchoring our broad strain continuum with the entirely doping-free NAO substrate, we unambiguously isolate biaxial compression as the intrinsic driving force for ambient-pressure superconductivity. This pristine phase diagram not only phenomenologically bridges the gap between ambient-pressure strained films and pressurized bulk crystals, but also establishes an ideal benchmark for future surface-sensitive probes like STM and ARPES to directly map delicate correlation effects and fundamentally resolve the multi-orbital pairing mechanism of nickelate superconductors.

In summary, by employing precision atomic-layer epitaxy to grow $La_2PrNi_2O_7$ films on diverse substrates, we explored a comprehensive strain continuum ranging from compressive (−2.14%) to tensile (+0.91%). For the first time, we achieved electron-like high-temperature superconductivity with a $T_{c,onset} \approx 60$ K and a Meissner state emerging below 20 K on Sr-free $NdAlO_3$ substrates, unambiguously decoupling intrinsic biaxial strain from interfacial doping. Furthermore, our results reveal the systematic evolution of lattice parameters and the Hall coefficient across a broad strain regime, spanning from compression to tension, in the bilayer nickelate system. Notably, a comparison with high-pressure bulk data supports that robust superconductivity emerges concomitant with the suppression of SDW, despite divergent lattice responses. By demonstrating that this superconducting state spans contrasting regimes—from electron-like films to hole-like bulk crystals—our work underscores the fundamental distinctions between epitaxial and pressurized systems. Ultimately, we establish a robust paradigm whereby in-plane biaxial strain modulates both Fermi surface topology and electronic correlations, providing critical empirical groundwork for theoretical models to capture the true essence of nickelate superconductivity.

## Main text figures

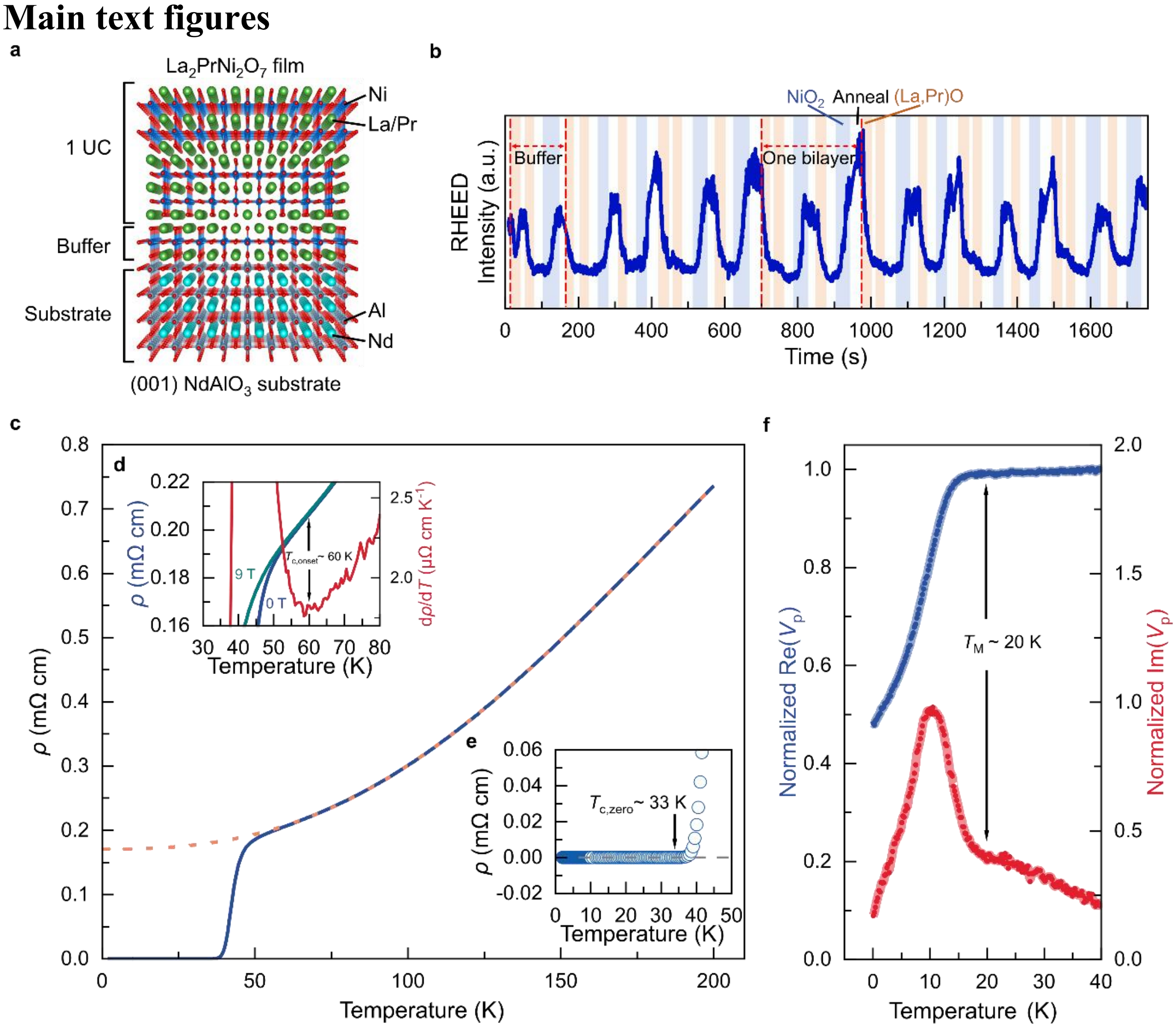

**Fig.1 | Superconducting $La_2PrNi_2O_7$ thin film on $NdAlO_3$ substrate. a,** The structural schematic of a $La_2PrNi_2O_7$ bilayer nickelate film grown on (001)-oriented $NdAlO_3$ substrate. **b**, RHEED oscillations of $La_2PrNi_2O_7$ growth on $NdAlO_3$. Orange and blue blocks represent the growth of $(La,Pr)O_x$ and $NiO_x$ layer, respectively. **c**, $\rho$–$T$ curve for a 3UC $La_2PrNi_2O_7$ film on $NdAlO_3$ (Sample A). MIR fitting of the normal-state resistivity over the 60–200 K range yields $\rho_0 = 0.188\, m\Omega \cdot cm$ , $A = 1.12 \times 10^{-6}\, m\Omega \cdot cm\, /\, K^{2.59}$ , $n = 2.59$ , $\rho_{sat} = 1.86\, m\Omega \cdot cm$ (see Methods), with the fitted curve plotted in orange. **d,** $\rho$–$T$ curve zoomed in near $T_{c,onset}$. The onset $T_c = 60$ K (black arrow) is defined as the temperature-dependent deviation of resistivity (red line) reach minimum and coincides with the onset of the 9 T magnetic field response (green line). **e**, $\rho$–$T$ curve zoomed in near the zero-resistance state. $T_{c,zero} = 33$ K is defined as the film resistance drops below the instrumental resolution limit. **f**, Normalized real part (Re(Vp)) and imaginary part (Im(Vp)) of pickup coil voltage as functions of temperature, measured using a two-coil mutual-inductance technique on Sample A.

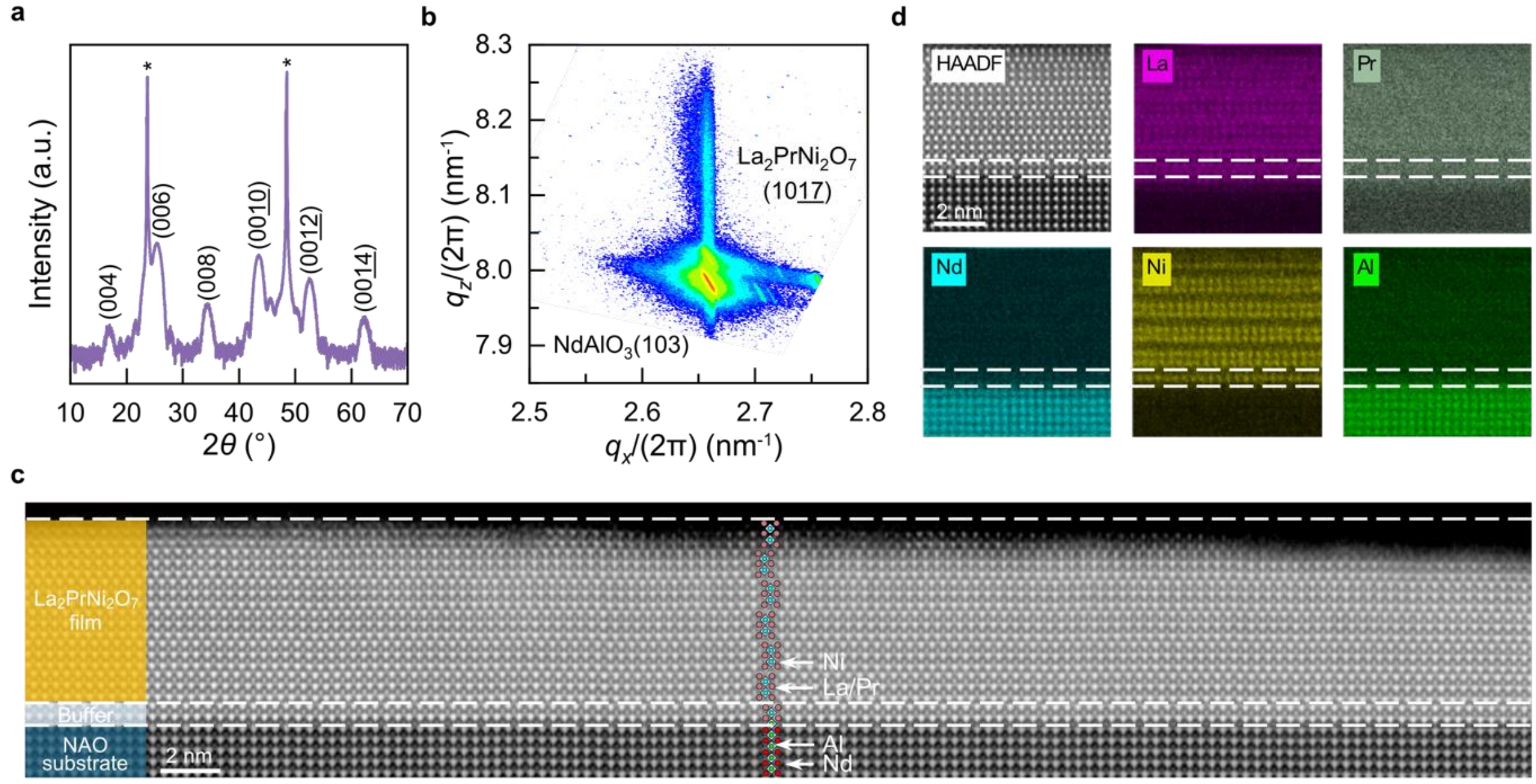

**Fig.2 | Epitaxial strain and atomic-scale structural characterization of a superconducting $La_2PrNi_2O_7$ film on $NdAlO_3$ substrate. a**, XRD $2\theta$–$\omega$ scans of a $La_2PrNi_2O_7$ film on $NdAlO_3$ substrate. **b**, Reciprocal space mapping around the $NdAlO_3$ (103) and $La_2PrNi_2O_7$ (10$\underline{17}$) reflections, confirming coherent epitaxial strain between the film and substrate. The additional substrate signal originates from substrate twinning. **c**, Large-scale high-angle annular dark-field scanning transmission electron microscopy (HAADF-STEM) images of a 3UC $La_2PrNi_2O_7$ film on $NdAlO_3$ substrate. Dashed lines represent the buffer/interface between the film and the NAO substrate. The loss of some outermost unit cells may originate from damage during FIB sample preparation. **d**, HAADF image and atomic-resolution EDS elemental maps (La, Pr, Nd, Ni, Al) of the same region. The area between the dashed lines indicates the artificially constructed buffer/interface.

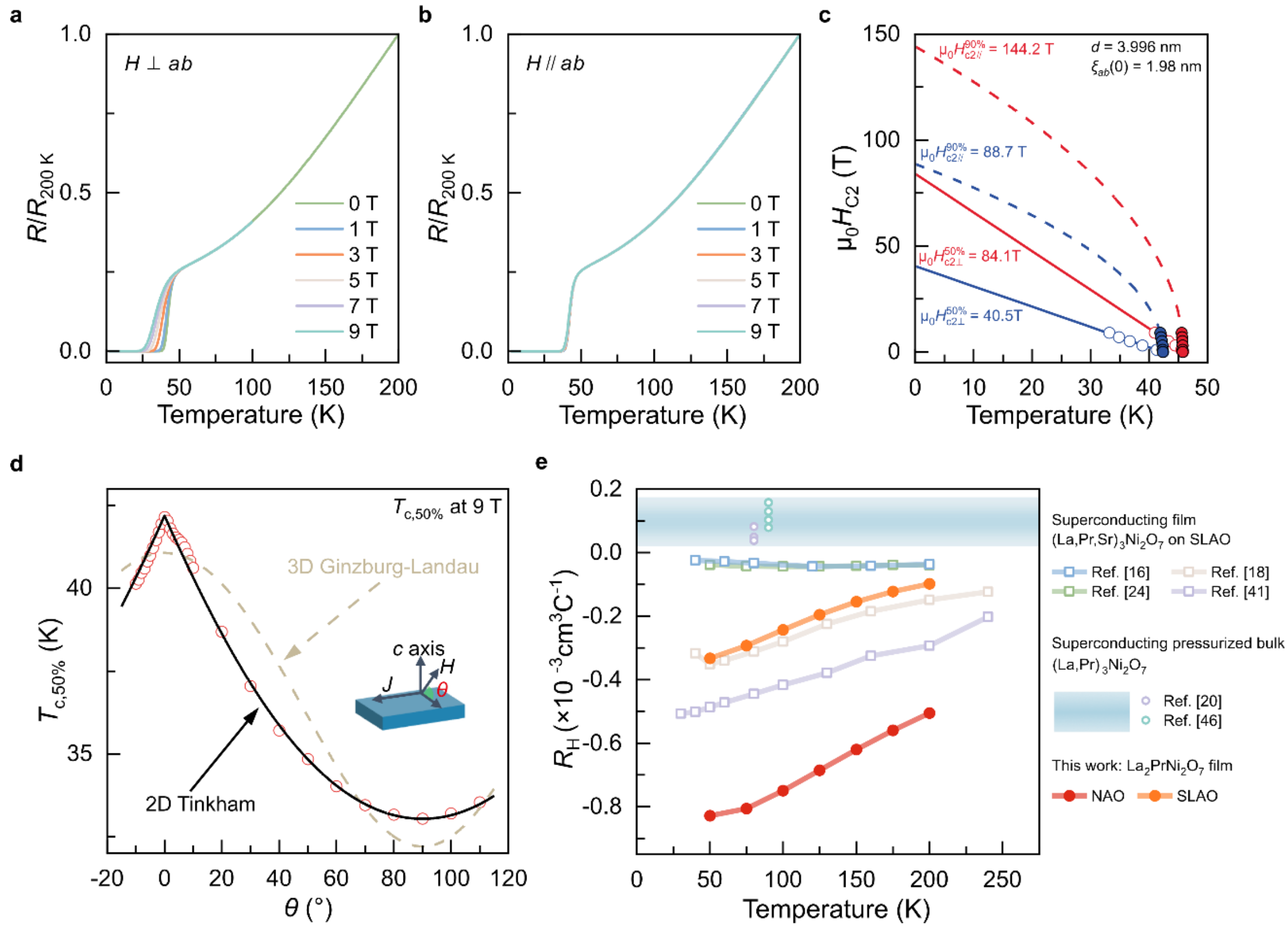

**Fig.3 | Quasi-2D superconductivity and Hall coefficient evolution in $La_2PrNi_2O_7$ film on $NdAlO_3$ substrate. a**,**b**, The $R/R_{200\,K}$ curves of a $La_2PrNi_2O_7$ thin film under various magnetic fields applied perpendicular (a) and parallel (b) to the *ab* plane of the film. **c**, Temperature dependence of the upper critical fields $H_{c2}$ for $H \parallel ab$ (solid circles) and $H \perp ab$ (open circles). Critical fields defined by 90% (red circles) and 50% (blue circles) of normal state resistance $R_N$ are presented. Solid lines are linear fits. Dashed lines are fits using the 2D Ginzburg–Landau formula. **d**, Angular dependence of $T_{c,50\%}$ at 9 T fitted with the 2D Tinkham model (solid line) and 3D Ginzburg–Landau model (dashed line). Open circles are the measured $T_{c,50\%}$ values. Inset shows the measurement geometry. **e**, Temperature dependent Hall coefficient ($R_H$) for $La_2PrNi_2O_7$ films and references (other optimized bilayer films and superconducting bulk near optimal pressure). The $R_H$ data for films are shown as symbols connected by lines, while the shaded region represents the range of values reported for bulk.

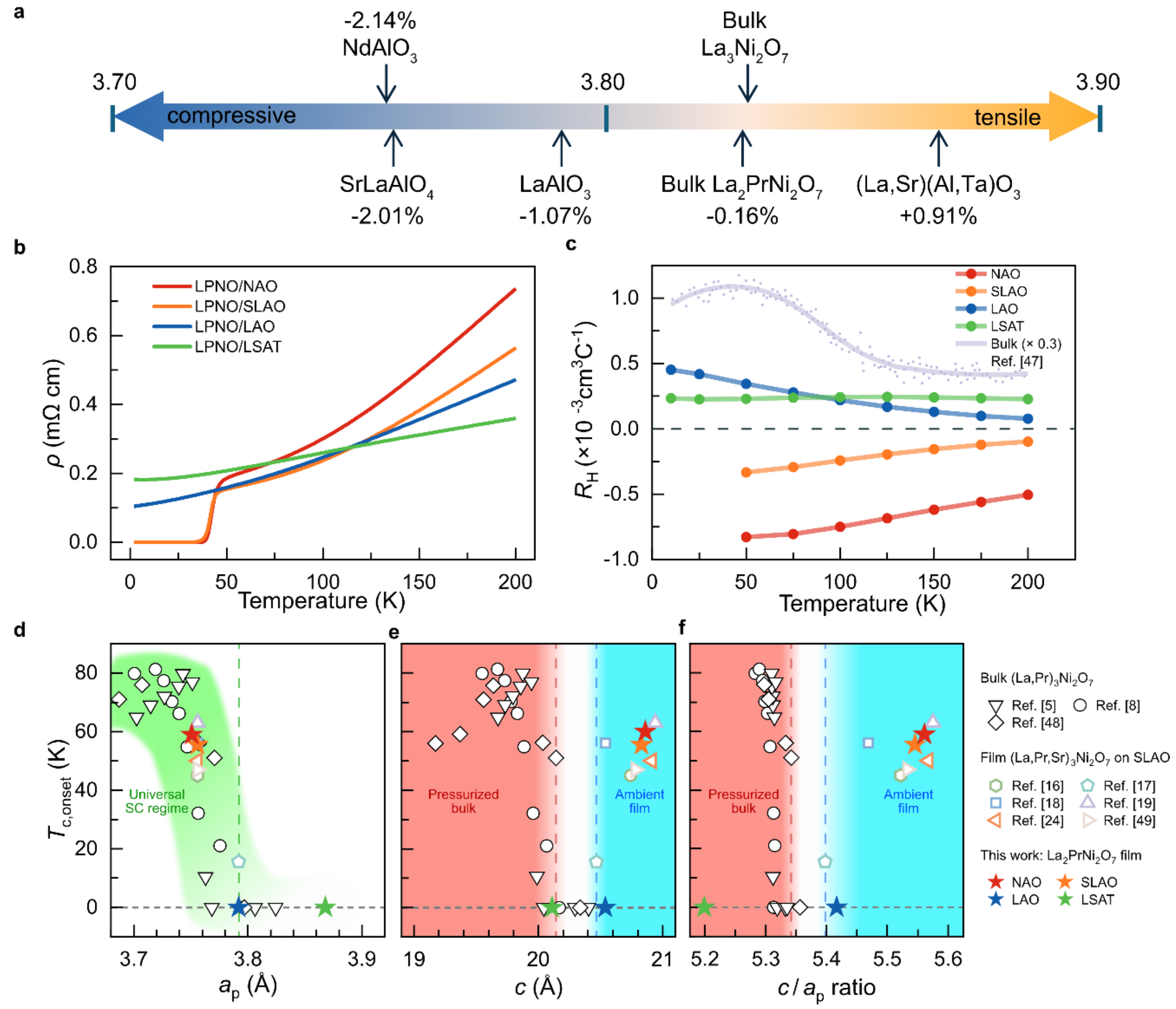

**Fig.4 | Strain-dependent structural and electronic properties among $La_2PrNi_2O_7$ thin films. a**, Epitaxial strain continuum comparing pseudo-cubic (pc) substrates with bulk pseudo-tetragonal (pt) $La_3Ni_2O_7$. Values in parentheses denote the corresponding strain coherent states of LPNO films. **b,** Representative $\rho$–$T$ curves are shown for films grown on NAO, SLAO, LAO and LSAT substrates, respectively. **c**, Temperature dependent $R_H$ of $La_2PrNi_2O_7$ films grown on NAO, SLAO, LAO and LSAT substrates. The bulk curve is multiplied by a factor of 0.3 for better visualization. **d,e,f**, Comparison of $T_{c,onset}$ versus in-plane lattice constant ($a_p$), out-of-plane lattice constant ($c$), and $c/a_p$ ratio. Open black and colored symbols denote pressurized bulk and film data from the literature, respectively. Each point is the highest $T_{c,onset}$ sample from each reference, using the same $T_{c,onset}$ definition as in Fig. 1d. The green shaded region indicates that superconductivity appears in both films and bulk materials upon reduction of the in-plane lattice parameter ($a_p$). The red and blue shaded regions indicate the distinct structural regimes where superconductivity emerges in pressurized bulk and ambient-pressure thin films, respectively.

## Methods

### Thin-film growth

By alternately ablating (La,Pr)$O_x$ and $NiO_x$ targets with different laser energies under in situ reflection high-energy electron diffraction (RHEED) monitoring, we kinetically suppressed stoichiometric deviations, thereby achieving atomic-layer-by-layer growth of films with exceptional crystalline quality in a highly oxidative environment. All samples were prepared on as-received $NdAlO_3$(001), $SrLaAlO_4$(001), $LaAlO_3$(001), and (La,Sr)(Al,Ta)$O_3$(001) substrates (MTI-Kejing) using our optimized energy-switchable, ozone-assisted atomic-layer-by-layer epitaxy technique. The laser energy was tunable via the software-controlled insertion of $Al_2O_3$ single crystals with calibrated thicknesses into the optical path.

The $La_{0.67}Pr_{0.33}O_x$ powders were synthesized via a standard sol-gel method from a stoichiometric mixture of $La(NO_3)_3{\cdot}6H_2O$ and $Pr(NO_3)_3{\cdot}6H_2O$. The resulting powders were pressed into pellets and sintered twice, each time for 12 h at 1150 °C in a flowing Ar atmosphere. The $NiO_x$ target was purchased from PrMat.

The ablation sequences for the perovskite substrates followed a stacking order of (La,Pr)O–(La,Pr)O–$NiO_2$–[(La,Pr)O–(La,Pr)O–$NiO_2$–(La,Pr)O–$NiO_2$] × n, while those for the $K_2NiF_4$-type substrates followed (La,Pr)O–$NiO_2$–[(La,Pr)O–(La,Pr)O–$NiO_2$–(La,Pr)O–$NiO_2$] × n. Film stoichiometry was controlled by precisely adjusting the number of laser pulses required to deposit a single atomic layer. Preliminary calibration of the pulse numbers for the two targets was performed by growing $(La,Pr)_2O_3$ on YSZ(111) and NiO on MgO(100) prior to growing the bilayer phase on the target substrates, following a calibration method previously reported for the molecular beam epitaxy (MBE) growth of nickelates[31,63,64]. Typically, 110-120 pulses were used to ablate the $La_{0.67}Pr_{0.33}O_x$ and $NiO_x$ targets, achieving a stoichiometric precision of better than 1%.

Film growth was carried out at a temperature of 750 °C under a mixed atmosphere of purified ozone and oxygen. The total growth pressure was maintained at 10 Pa, with an ozone partial pressure of approximately 2.1 Pa. To minimize ozone decomposition, we introduced ozone into the chamber through a water-cooled nozzle located 3 cm from the sample holder. The laser fluences were set to 1.6 and 2.0 J/cm$^2$ for the (La,Pr)$O_x$ and $NiO_x$ targets, respectively, with a pulse repetition rate of 4 Hz. Following deposition, the samples were cooled at a rate of 60 °C/min to below 160 °C, transferred to the load-lock chamber, and then immediately backfilled with pure oxygen to prevent oxygen loss.

### Structural characterization and transport measurements

The structural properties of the films were characterized by X-ray diffraction (XRD) and reciprocal space mapping (RSM) using Cu K$\alpha_1$ ($\lambda$ = 1.5406 Å) radiation (Rigaku SmartLab). Microstructural and elemental analyses of the thin films were conducted via high-angle annular dark-field scanning transmission electron microscopy (HAADF-STEM), annular bright-field (ABF) and energy-dispersive X-ray spectroscopy (EDS) on a JEOL JEM-ARM300F microscope integrated with a probe-forming spherical aberration (Cs) corrector and JEOL-built dual silicon drift detectors (SSD).

Electrical transport properties were measured using the standard Hall-bar geometry and the van der Pauw method in a Physical Property Measurement System (PPMS; Quantum Design). Pt electrodes were patterned by magnetron sputtering and connected via aluminum wire bonding.

### Mutual inductance measurements

The Meissner state was measured using a two-coil mutual inductance setup adapted to a dilution refrigerator system. Both coils have 300 turns, are 1.5 mm long, and are spaced ~1 mm apart on a sapphire block fixed with G10 epoxy. During the measurement, a 0.1 mA AC current at 23.37 kHz was applied to the drive coil, and the mutual inductance was recorded using an SR830 lock-in amplifier.

### Upper critical field and dimensionality analysis

The upper critical fields $H_{c2}(T)$ were extracted using both 90% and 50% criteria of the extrapolated normal-state resistance $R_N$. For the out-of-plane field ($H \perp ab$), the critical field exhibits a typical linear temperature dependence and was fitted using the standard linearized Ginzburg-Landau (GL) model:

$$H_{c2}^{\perp ab}(\mathrm{T}) = H_{c2}^{\perp ab}(0)\,(1 - \frac{T}{T_c})$$

In contrast, the in-plane critical field ($H \parallel ab$) was evaluated using the 2D GL formula for a geometrically confined superconductor, which accounts for the non-linear square-root dependence:

$$H_{\mathrm{c2}}^{\parallel ab}(\mathrm{T}) = H_{\mathrm{c2}}^{\parallel ab}(0)\sqrt{1-\frac{T}{T_{\mathrm{c}}}}$$

Assuming the 90% criterion represents the intrinsic upper critical field, the zero-temperature in-plane coherence length $\xi_{ab}(0)$ and the effective superconducting layer thickness $d$ were calculated via $H_{\mathrm{c2}}^{\perp ab}(0) = \phi_0/2\pi\xi_{ab}^2(0)$ and $H_{\mathrm{c2}}^{\parallel ab}(0) = \sqrt{12}\phi_0/2\pi\xi_{ab}(0)d$, respectively, where $\phi_0$ is the magnetic flux quantum.

For the angular dependence of the transition temperature $T_{\mathrm{c}}(\theta)$ under a contrast magnetic field $H$, the strictly 2D behavior was verified using the 2D Tinkham model:

$$T_{\mathrm{c}}(H,\theta) = T_{\mathrm{c,zero}} - \left|\left(T_{\mathrm{c,zero}} - T_{\mathrm{c}}^{H\perp ab}(H)\right)\sin\theta\right| - \left(T_{\mathrm{c,zero}} - T_{\mathrm{c}}^{H\parallel ab}(H)\right)\cos^2\theta$$

where $T_{\mathrm{c,zero}}$ is the zero-field transition temperature. Conversely, the data were tested against the 3D anisotropic GL model:

$$T_{\mathrm{c}}(H,\theta) = T_{\mathrm{c,zero}} + \frac{H}{\mathrm{d}T_{\mathrm{c}}^{H\perp ab}(H)/\mathrm{d}T}\left(\sin^2\theta + \gamma^{-2}\cos^2\theta\right)^{1/2}$$

where $\gamma$ represents the effective mass anisotropy ratio. The experimental sharp cusp near $\theta = 0°$ conforms exclusively to the 2D Tinkham limit.

**Mott-Ioffe-Regel (MIR) limit fitting**

The high-temperature saturation of the resistivity was analyzed using the parallel-resistor model, which assumes that the measured resistivity $\rho(T)$ is a parallel combination of the ideal temperature-dependent resistivity $\rho_{ideal}(T)$ and a temperature-independent saturation resistivity $\rho_{sat}(T)$ dictated by the MIR limit. The data were fitted using the following equation:

$$\frac{1}{\rho(T)} = \frac{1}{\rho_{ideal}(T)} + \frac{1}{\rho_{\mathrm{sat}}}$$

where $\rho_{ideal}(T)$ is described by $\rho_{ideal}(T) = \rho_0 + AT^n$. Here, $\rho_0$ represents the residual resistivity, $A$ is the scattering coefficient, and the exponent $n$ reflects the dominant scattering mechanism in the $La_2PrNi_2O_7$ films. $\rho_{\mathrm{sat}}$ corresponds to the maximum metallic resistivity when the mean free path reaches the order of the lattice constant. All fitting parameters were extracted using a non-linear least-squares minimization algorithm.

## Extended Figures and Tables

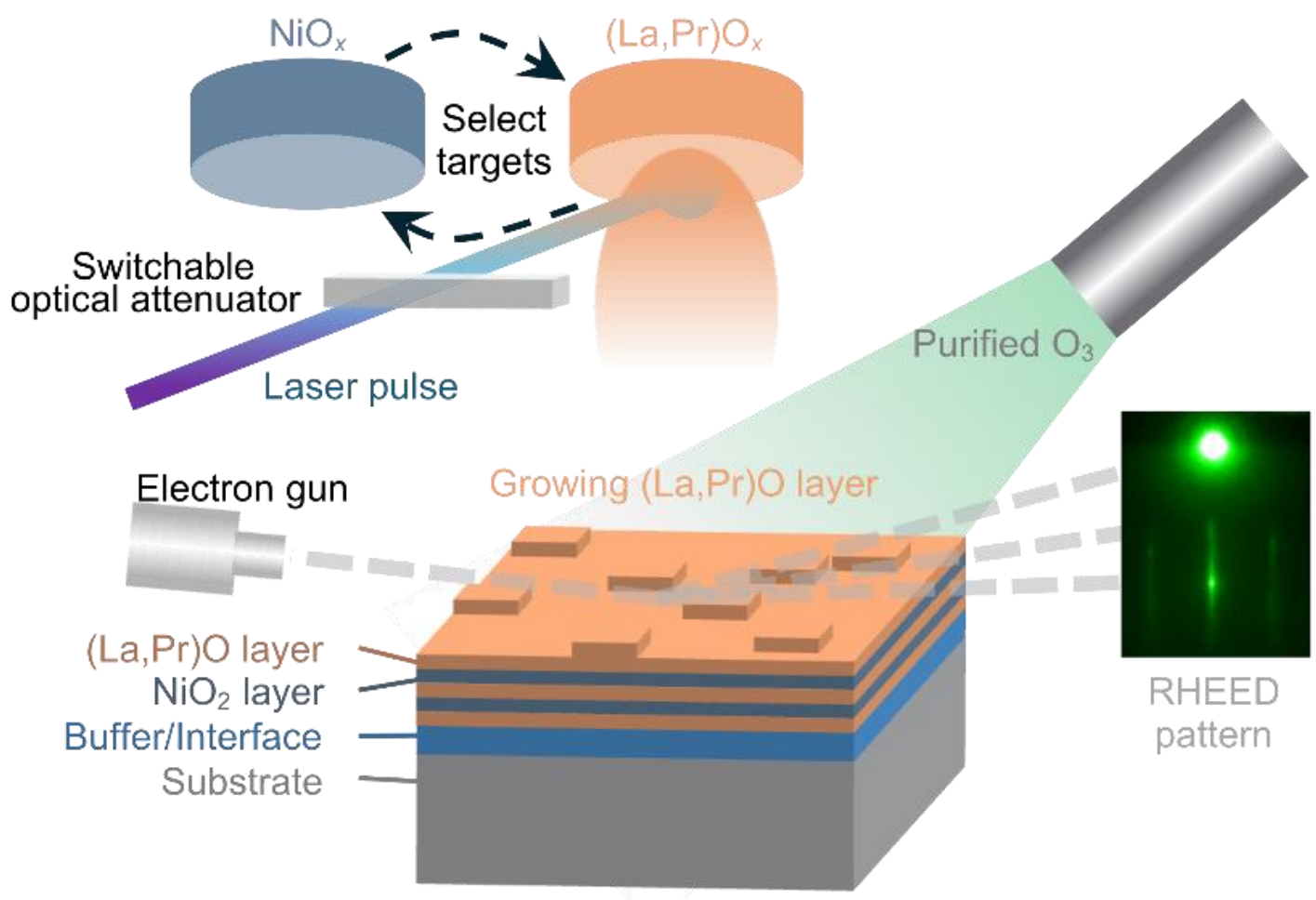

**Extended Data Fig. 1 | Schematic of energy-switchable ozone-assisted atomic-layer-by-layer epitaxy technique.** Through programmed switching of the laser energy and ablation targets, with real-time monitoring by double-differential RHEED, we achieve precise layer-by-layer control of (La,Pr)O and $NiO_2$ deposition under highly oxidizing conditions.

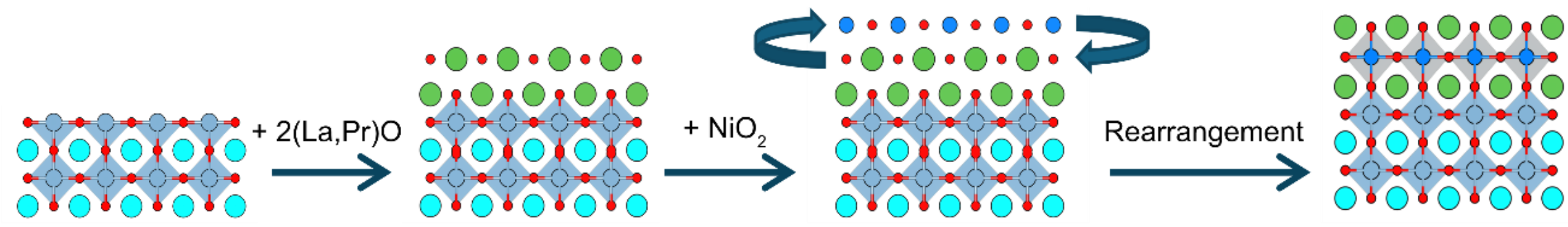

**Extended Data Fig. 2 | Schematic of dynamic atomic-layer rearrangement.** After two (La,Pr)O layers are deposited on the $AlO_2$-terminated substrate, the $NiO_2$ layer swaps with the outer (La,Pr)O layer. This behavior is not confined to the interfacial growth shown here; it persists throughout the entire atomic-layer-by-layer growth process.

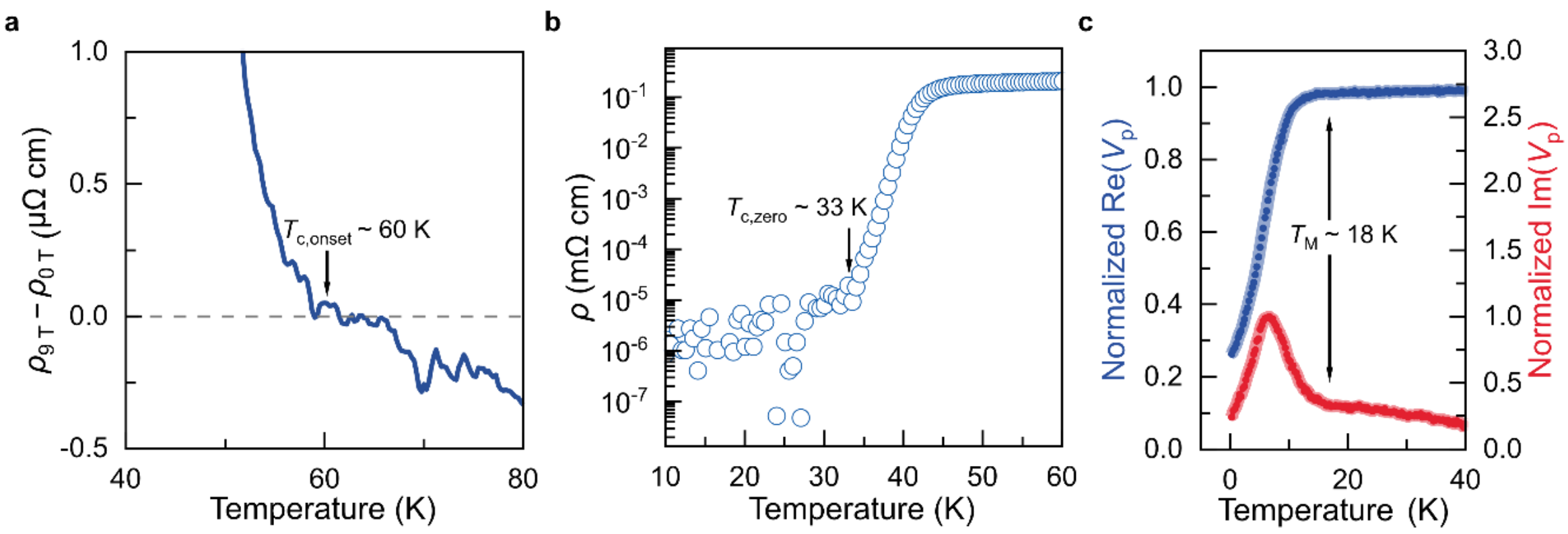

**Extended Data Fig. 3 | Extended transport characterizations of superconducting $La_2PrNi_2O_7$/NAO films. a,** Temperature dependence of the field-induced resistivity change for the Sample A shown in Fig. 1d. **b,** The *ρ*–*T* curve (Sample A) corresponding to Fig. 2c, plotted on a semi-logarithmic scale **c,** Two-coil mutual inductance measurement of Sample B, which exhibits transport properties similar to those of Sample A.

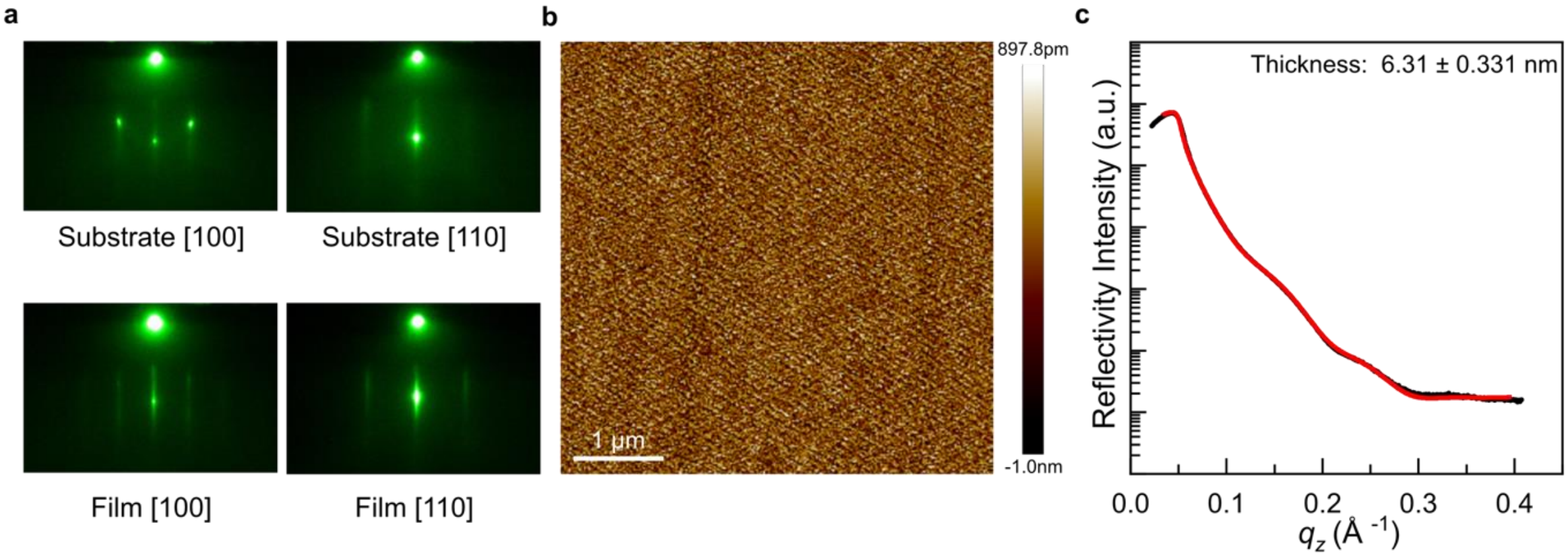

**Extended Data Fig. 4 | Surface and thickness of a superconducting $La_2PrNi_2O_7$/NAO film. a**, Reflection high-energy electron diffraction (RHHED) patterns of substrate and thin film taken along [100] and [110] directions. **b**, AFM image of a 3UC LPNO film on NAO substrate. The root-mean-square roughness is 279 pm. **c**, X-ray reflectivity (XRR) of 3UC LPNO/NAO film.

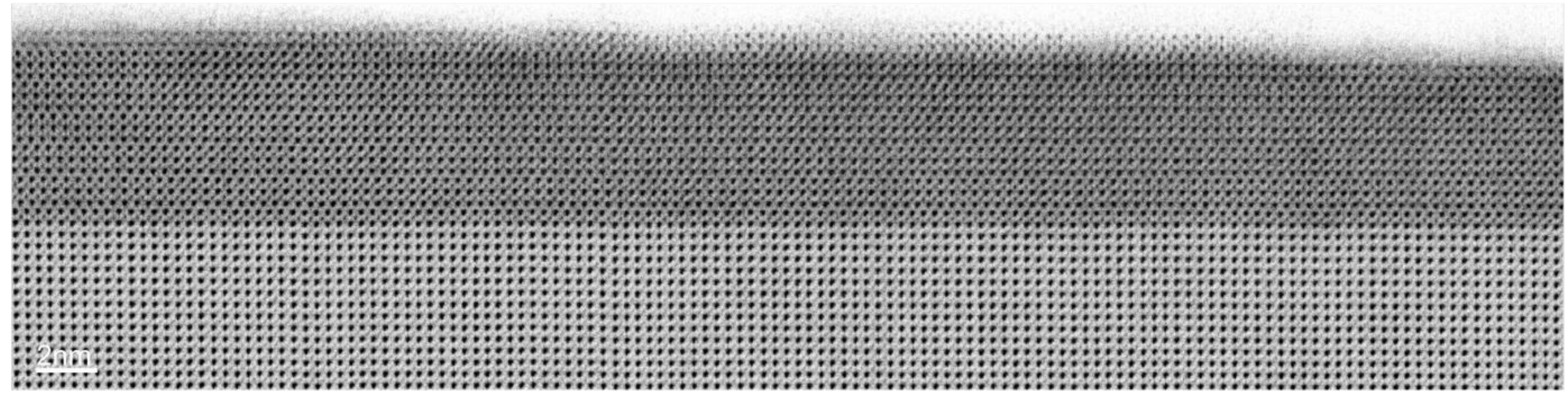

**Extended Data Fig. 5 | Large-scale annular bright-field (ABF) image of the $La_2PrNi_2O_7$ film.** The image exhibits a structural profile consistent with that shown in Fig. 2c.

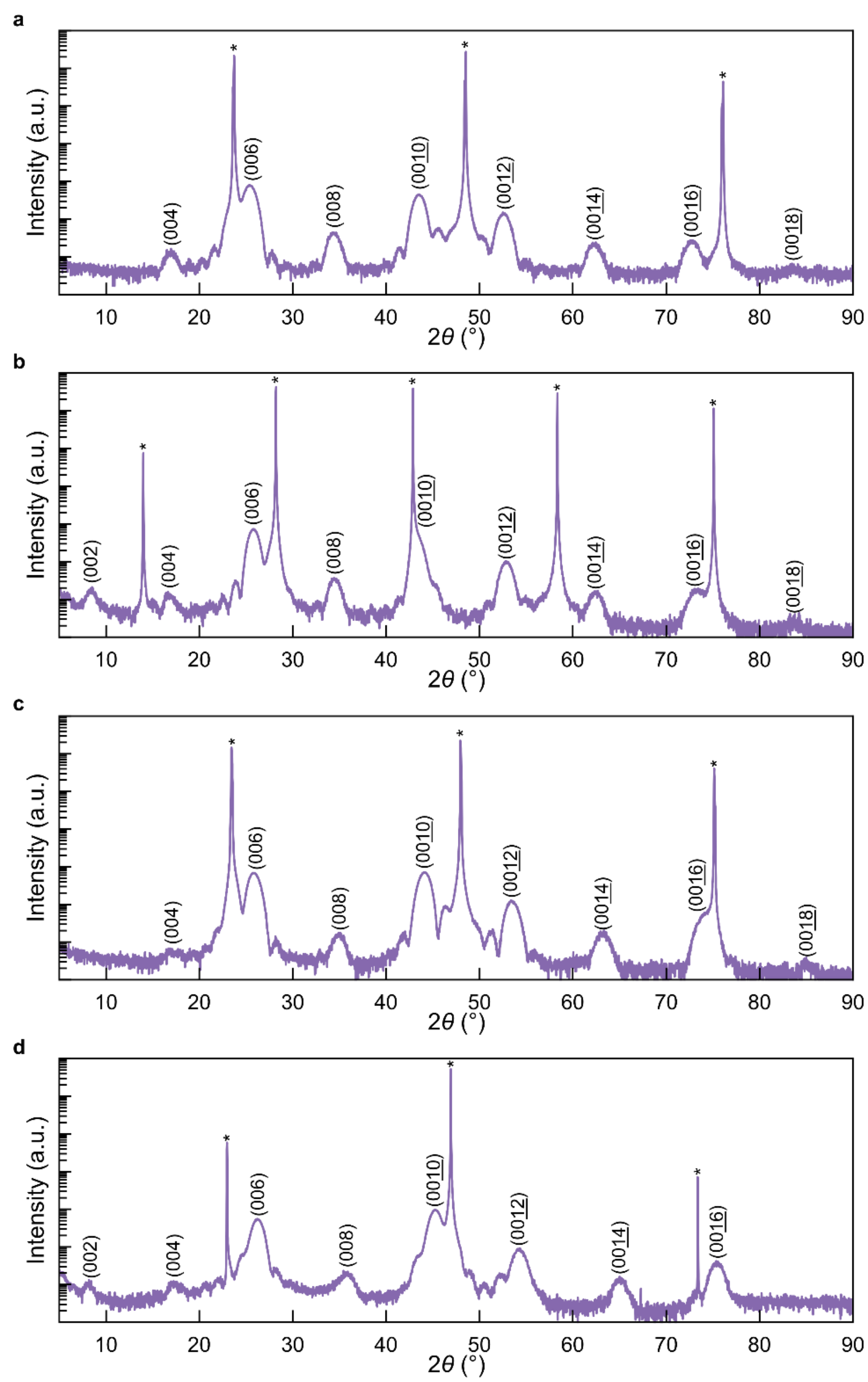

**Extended Data Fig. 6 | Wide-range 2*θ*–*ω* XRD scans of the 3UC $La_2PrNi_2O_7$ films. a – d,** Scans are shown for films grown on **a,** NAO, **b,** SLAO, **c,** LAO and **d,** LSAT substrates, respectively. The sharp (00*l*) diffraction peaks and pronounced Laue fringes confirm the exceptional crystalline quality of all films.

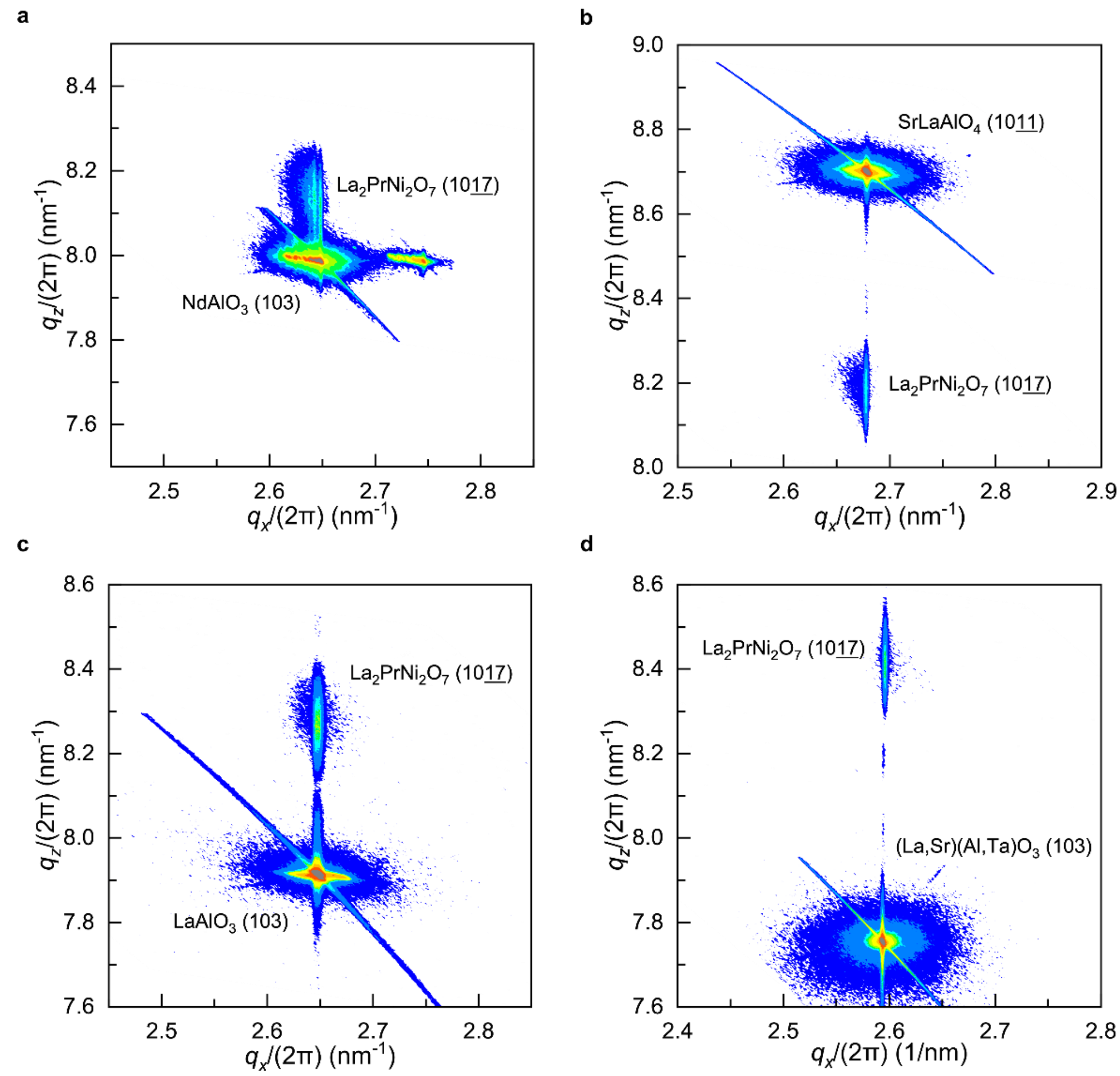

**Extended Data Fig. 7 | Reciprocal space maps of the 3UC $La_2PrNi_2O_7$ films. a – d,** RSMs are shown for films grown on **a,** NAO, **b,** SLAO, **c,** LAO and **d,** LSAT substrates, respectively. In a, the additional reflections from both the substrate and the film originate from substrate twinning. Collectively, these maps confirm highly coherent epitaxial growth across the entire strain continuum.

**Extended Data Table 1 | $La_2PrNi_2O_7$ films experimental structural parameters.** Substrate lattice parameters for all substrates were derived from supplier data, whereas the pseudo-cubic parameters of the recently commercialized NAO substrate were obtained via four-circle diffraction. Nominal strain calculated from lattice parameters as following $((a_{sub,pc} - a_{bulk,pt}) / a_{bulk,pt}$ where $a_{bulk,pt}$ = 3.833 Å; film $c$ lattice parameters and errors fit from Nelson-Riley fits to $2\theta$–$\omega$ scans.

| | $NdAlO_3$ | $SrLaAlO_4$ | $LaAlO_3$ | $(La,Sr)(Al,Ta)O_3$ | Bulk[5] |
|---|---|---|---|---|---|
| pseudo-cubic *a*, *b* axis (Å) | 3.751 | 3.756 | 3.792 | 3.868 | 3.833 |
| nominal strain (%) | -2.14 | -2.01 | -1.07 | 0.91 | - |
| *c* lattice axis (Å) | 20.86 ± 0.09 | 20.83 ± 0.03 | 20.54 ± 0.05 | 20.11 ± 0.04 | 20.52 |
| orthorhombic unit cell volume ($Å^3$) | 587.0 | 587.7 | 590.7 | 601.7 | 603.0 |

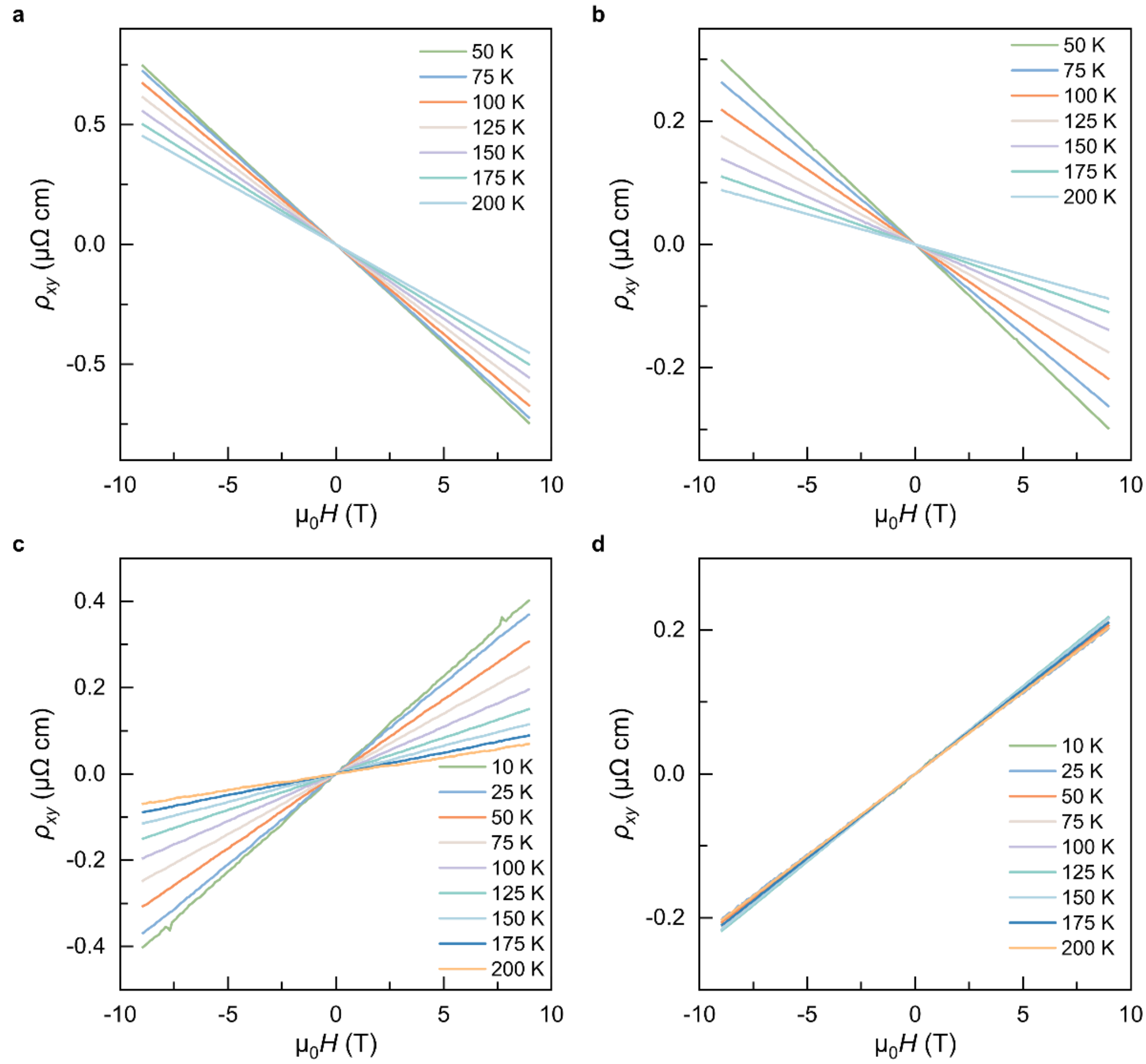

**Extended Data Fig. 8 | Hall resistivity of 3UC $La_2PrNi_2O_7$ films. a – d,** Hall resistivity is shown for films grown on **a,** NAO, **b,** SLAO, **c,** LAO and **d,** LSAT substrates, respectively. All films exhibit linear magnetic-field dependence across the full range from −9 T to 9 T.

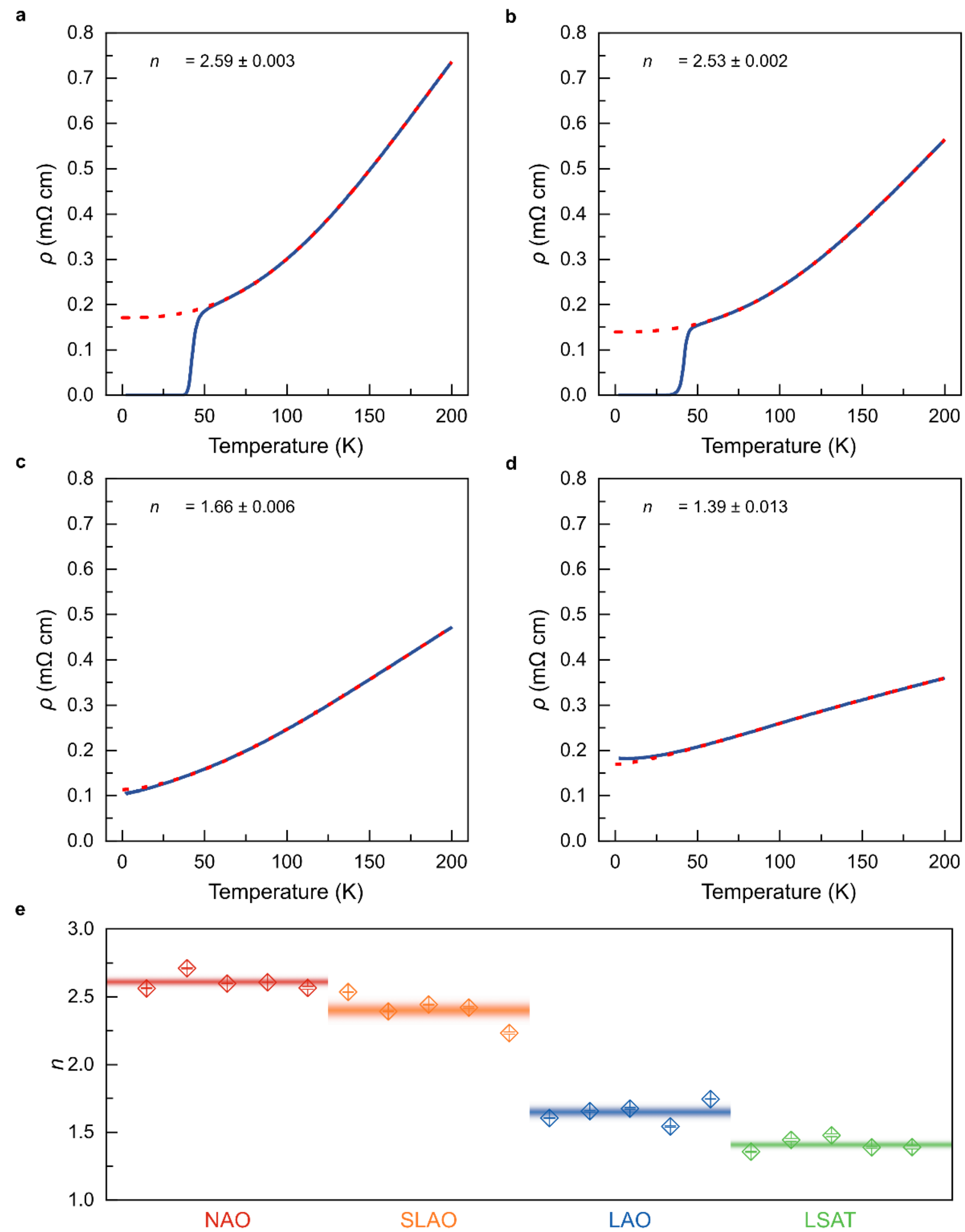

**Extended Data Fig. 9 | Transport and MIR fitting analysis of the $La_2PrNi_2O_7$ films. a – d,** Representative transport data and the corresponding MIR fits are shown for films grown on **a,** NAO, **b,** SLAO, **c,** LAO and **d,** LSAT substrates, respectively. **e,** Comparison of $n$ for LPNO/NAO (red), LPNO/SLAO (orange), LPNO/LAO (blue) and LPNO/LSAT (green), respectively. All error bars represent the standard deviations of the parameters obtained from the fitting. The shaded horizontal bands indicate the average values of each parameter, with their widths corresponding to twice the standard deviation derived from statistics over five samples.

**Extended Data Table 2 |** Average values of fit parameters $n$ for LPNO/NAO, LPNO/SLAO, LPNO/LAO and LPNO/LSAT. Values are averages over five samples and uncertainties are the standard deviations.

| | LPNO/NAO | LPNO/SLAO | LPNO/LAO | LPNO/LSAT |
|---|---|---|---|---|
| $n$ | 2.61 ± 0.06 | 2.40 ± 0.11 | 1.65 ± 0.08 | 1.41 ± 0.05 |

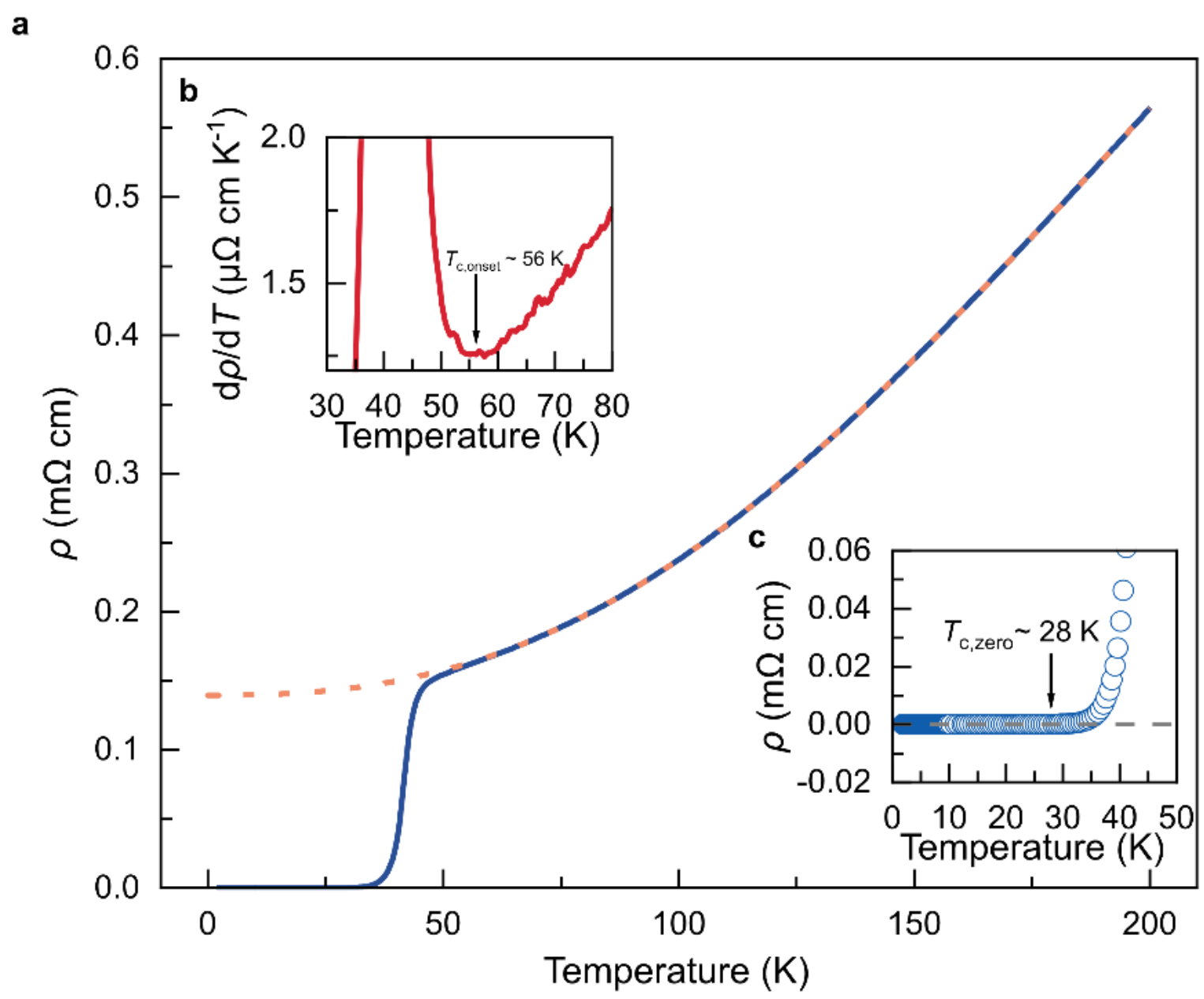

**Extended Data Fig. 10 | Superconducting $La_2PrNi_2O_7$ thin film with $SrLaAlO_4$ substrate. a**, *ρ–T* curve for a 3UC $La_2PrNi_2O_7$ film on $SrLaAlO_4$. MIR fitting of the normal-state resistivity over the 60–200 K range yields $\rho_0 = 0.153\ m\Omega \cdot cm$, $A = 1.09 \times 10^{-6}\ m\Omega \cdot cm\ /\ K^{2.54}$, $n = 2.54$, $\rho_{sat} = 1.52\ m\Omega \cdot cm$, with the fitted curve plotted in orange. **b,** *ρ–T* curve zoomed in near $T_{c,\mathrm{onset}}$. The onset $T_c$ = 56 K (black arrow) is defined as the temperature-dependent deviation of resistivity (red line) reach minimum. **c**, *ρ–T* curve zoomed in near the zero-resistance state. $T_{c,\mathrm{zero}}$ = 28 K is defined as the film resistance drops below the instrumental resolution limit.

## Acknowledgements

We are thankful for the valuable discussion with Junfeng He, Kun Jiang and the assistance of XRD measurement from Wei Hu. This work is supported by the National Natural Science Foundation of China (Grant Nos. 12488201, 12494592, 12325403), the National Key R&D Program of the MOST of China (Grant No. 2022YFA1602601), the Chinese Academy of Sciences under contract No. JZHKYPT-2021-08, the CAS Project for Young Scientists in Basic Research (Grant No. YBR-048), the CAS Superconducting Research Project under Grant No. [SCZX-0101] and the Innovation Program for Quantum Science and Technology (Grant No. 2021ZD0302800). This work was partially carried out at Instruments Center for Physical Science, University of Science and Technology of China.

**Author Contributions**

X.H.C. conceived the research project and coordinated the experiments. Z.W.W. performed thin-film growth with assistance from Z.Y.H. Z.W.W. and Z.J.W. performed low-temperature measurements and analysis. H.Y.W., M.Y.Z., S.C.N. and J.T. contributed on STEM measurement and analysis. M.Z.S and H.P.L contributed on XRD measurement and analysis. Z.W.W., T.W. and X.H.C. interpreted the results and wrote the manuscript. All authors discussed the results and commented on the manuscript.

**Competing Interests**

The authors declare no competing interests.